# Collaborative Framework with Shared Responsibility for Relief Management in Disaster Scenarios

*Bhupesh Kumar Mishra[1], Keshav Dahal[2]*

*1: bhupeshmishra@hull.ac.uk, Data Science, AI & Modelling Centre (DAIM), University of Hull, UK*

*2: keshav.dahal@uws.ac.uk, School of Engineering, computing and Physical Science, the University of the West of Scotland, UK*

**Abstract**

Disasters instances have been increasing both in frequency and intensity causing the tragic loss of life and making life harder for the survivors. Disaster relief management plays a crucial role in enhancing the lifestyle of disaster victims by managing the disaster impacts. Disaster relief management is a process with many collaborative sectors where different stakeholders should operate in all major phases of the disaster management progression. In the different phases of the disaster management process, many collaborative government organisations along with nongovernment organisations, leadership, community and media at different levels need to share the responsibility with disaster victims to achieve effective disaster relief management. Shared responsibility enhances disaster relief management effectiveness and reduces the disaster's impact on the victims. Considering the diverse roles of different stakeholders, there has been a need for a framework that can bind different stakeholders together during disaster management. this paper shows a framework with major stakeholders of disaster relief management and how different stakeholders can take part in an effective disaster relief management process. The framework also highlights how each stakeholder can contribute to relief management at different phases after a disaster. The paper also explores some of the shared responsibility collaborative practices that have been implemented around the world in response to the disaster as a disaster relief management process. In addition, the paper highlights the knowledge obtained from those disaster instances and how this knowledge can be transferred and can be helpful in disaster mitigation and preparedness for future disaster scenarios.

**Keywords: Disaster; Relief Management; Collaborative Framework; Shared Responsibility**

## 1. Introduction

A disaster is defined as "a serious disruption of the functioning of a community or a society causing widespread human, material, economic or environmental losses which exceed the ability of the affected community or society to cope using its resources" (UNISDR: United Nations International Strategy for Disaster Reduction). Every year, natural disasters, such as earthquakes, hurricanes, landslides, volcanic eruptions, tsunamis, avalanches, extreme colds, heatwaves, and cyclones kill thousands of people around the world. The nature and level of disasters vary from small to large scale and affect people's lives. Therefore, the need for a quick response with relief items is crucial. After a disaster, a sufficient amount of relief items must be distributed over some time to decrease human losses and improve the quality of life of the survivors. To achieve this, an effective disaster relief management framework is crucial to distribute relief items such as food, water, medical supplies, and clothing to the disaster-affected areas which enhances the quality of life of disaster victims.

Disasters often occur without caution and unexpectedly leading to a series of adverse consequences. Therefore, disaster relief management must always be prepared to address these consequences effectively. United Nations defined disaster management as **"**The systematic process of using administrative decisions, organization, operational skills and capacities to implement policies, strategies and coping capacities of the society and communities to lessen the impacts of natural hazards and related environmental and technological disasters. This comprises activities, including structural and non-structural measures to avoid or to limit adverse effects of hazards" (Reduction, 2004). According to the Federal Emergency Management Agency (FEMA), the disaster management life cycle has been divided into four main phases: mitigation, preparedness, response, and recovery (Hoyos et al., 2015).

Each phase is crucial in making disaster relief management highly effective. Mitigation, rarely seen as urgent, is the effort to curtail the loss of life and property by minimising the impact of disasters (Bosher et al., 2007). The mitigation is achieved through risk analysis with available information. The risk analysis provides a foundation for mitigation activities for risk reduction. Preparedness within the field of disaster relief management is defined as a way of being ready to respond to a disaster, crisis or any other type of emergency (Guerdan, 2009). This phase includes developing plans for the actions that will improve the chances of successfully dealing with a disaster. The response phase deals with how to respond to the disaster. By being able to act responsibly and safely, the chance of survival increases in a disaster(Rawls and Turnquist, 2010). The recovery phase deals with a disaster where any actions must take care of the victims (Coles and Buckle, 2004). It requires the delivery of food, medicine, tents, sanitation equipment, tools and other necessities to people in distress, often for extended periods.

Over the years, several disaster relief management models have been used for the relief items distribution in disaster environments. However, there have been many disaster instances where the management had gone through challenges in terms of sharing responsibility effectively during the relief items distribution as disaster relief management has appeared as a complex task often associated with uncertainty and dynamics (Crosweller and Tschakert, 2021). Global disaster relief management initiatives have changed over the years and that has been continuously expanding and growing in terms of engagement of different stakeholders with a collective approach whereby every stakeholder has a role in disaster response (Lassa, 2018). In recent years, shared responsibility among the stakeholders such as government organisations, non-government organisations, leadership, citizens and media appeared to effective way of disaster relief management (Medel et al., 2020). In different disaster scenarios, the shared responsibilities have already proven very effective in saving lives and property and are dignified to remain to do so for years to come. Studies on disasters in Australia, China, the Philippines, Korea, Japan and Indonesia have shown that the accomplishment of effective disaster management depends on collaborative effort from multi-stakeholders along with the governmental policies, plans, and tools (Lin et al., 2017).

Studies have suggested that any disaster requires different stakeholders' involvement in disaster relief management with better integrated and cohesive interactions (Atkinson and Curnin, 2020). In this regard, a collaborative approach and initiative are required to be explored for disaster relief management that can assure cooperation among different major stakeholders. Studies have also presented that government collaboration with other stakeholders improves the efficiency of disaster relief management along with collaborative performance (Medel et al., 2020). These all researches have highlighted that there has been diverse organisational engagement and their role. Without collaboration, the effectiveness of relief management is always been compromised. These researches have also highlighted that in many of the scenarios, there has not been a guided framework to establish collaborative working. Sharing the responsibilities with well-established and controlled communication has been the need to achieve a higher level of effectiveness in disaster relief management. In this paper, we presented a collaborative framework with shared responsibility in disaster relief management. The framework has five key stakeholders, Government Organisations, non-governmental organisations (NGOs), Leadership, Community and Media, with shared responsibility for disaster relief management. Collaborative shared responsibility entails the engagement of key stakeholders of the governmental system and their obligations in disaster relief management. The framework will help to guide the roles and responsibilities of all the major stakeholders. One of the biggest advantages of collaborative shared responsibility is the use of knowledge resources from multiple sources as disasters always stretch the bounds of emergency resources and governmental preparedness. The collaborative shared responsibility approach gives the flexibility to bring all the stakeholders together where all the stakeholders contribute with their potential in the best way for disaster resilience.

The remainder of this paper is organised into the following sections. In section 2, a collaborative framework with shared responsibility for disaster relief management is presented which describes the framework architecture and components. In the section major roles of five key stakeholders in disaster, scenarios have been presented. This section also highlights how each stakeholder can contribute to the common goal to enhance the disaster victim's life. Finally, in the section, a discussion of the framework along with the conclusion of the paper is presented.

## 2. Collaborative Framework with Shared Responsibility

With the augmented scope and severity of disasters in recent years, it has become imperative to review the traditional disaster relief management approach characterized by a centralised government organisation. Though centralised government policies favour more controlled management for better and timely disaster relief management, a collaborative shared responsibility framework enhances the effectiveness of disaster resilience.

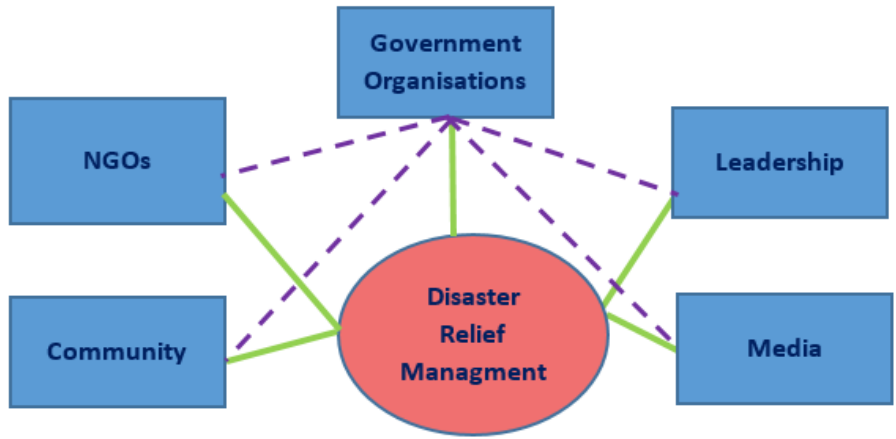

Fig. 1: Collaborative framework with five core elements having shared responsibility in disaster relief management

In general, collaboration is defined as any joint activity between two or more organisations that work together to achieve a common goal. Collaboration occurs when individuals/stakeholders from different organisations act together through their combined strength, resources, knowledge and skills and share ownership of the collaborations. Different types of collaboration can be formed across stakeholders with different levels of commitment for a specified period. When managing disaster relief distribution, the involvement of different organisations from different sectors needs to be coordinated effectively. This, however, requires comprehensive intra-organizational communications to maximise the capacity of organisations to handle a disaster impact.

A collaborative framework with five major stakeholders with shared responsibility for disaster relief management has been presented as shown in Figure 1. The collaborative framework incorporates five major stakeholders: Government Organisations, NGOs, Leadership, Community and Media. In the framework, the government organisations have been at the top of the hierarchy whereas the other four stakeholders are considered at the second level of the hierarchy. Disaster relief management tasks have been connected (green solid lines) with all five stakeholders, which also indicates the communication, information and activity sharing with the relief distribution process. The framework has also shown a connection (dotted purple lines) between government organisations to all other stakeholders. These lines show the control mechanism from the government organisations to all the stakeholders regarding disaster relief management. The control line reflects the obligation set by the government organisations to the corresponding stakeholders. The collaborative framework supports all basic disaster relief management factors under disaster situations. These factors are largely the system under which government organisations and other stakeholders operate and accomplish their tasks. These factors, namely the way organisations notice the disaster scenarios, the severity and its necessities, and the way they control and perform relief distribution tasks, are expected to affect the effectiveness of disaster resilience. These factors combined create a collaborative and shared responsible environment that determines the success or the failure of relief operations in terms of decision-making for disaster relief management.

## 3. Stakeholders of Collaborative Shared Responsibility for Disaster Relief Management

In the presented framework, the following five major stakeholders, and their responsibility in disaster relief management is discussed. The framework has been developed after reviewing several disaster relief management strategies implemented in different countries across different countries. Major stakeholders are identified based on the case-based scenarios presented in different disaster relief management case studies.

**3.1 Government Organisations:** Government organisations act as a system during disaster relief management with wealth, knowledge, resources, control and the power of decision-making (Crosweller and Tschakert, 2021). The government organisation has an abundance of valuable resources in terms of skilled professionals, relief items, money and other assets that can be managed to minimise the disaster's impact on the victims. A set of principles, processes and control mechanisms are pre-defined by the government organisation for disaster resilience. Government organisations can potentially accomplish their goal by sharing responsibility with other stakeholders and can play the controlling body to achieve their objectives (Busch and Givens, 2013). This controlled shared responsibility has been observed in the case of Hurricane Katrina in 2005 where Walmart played a crucial role in distributing relief items. A similar set of acts has been observed in the case of the Hebei Spirit oil spill in Korea where the government organisation has the advantage of disaster management in terms of centralised precise management (Cheong, 2011). This study has also shown how government organisations play the controlling body in the management of different geographical and cultural diversity.

**3.2 Non-Government Organisation:** Non-government Organisations (NGOs) such as the Red Cross, the United Nations, and the private sectors appeared more as service providers to individuals or communities in need during any disaster relief management. Their role can be observed as a substitute for government organisations' objectives and functions. There have been examples of NGOs' involvement with success stories around the world in many disaster scenarios such as Indian Ocean Tsunamis, and Japan Tsunamis (Lassa, 2018). In these disaster cases and others, the major roles of NGOs have been providing relief items, establishing health camps, engaging in rescue tasks, setting temporary shades and so on. As a success story, a study from the Bam earthquake, in Iran, showed that the government organisations and NGOs spontaneously engaged in disaster relief management in terms of rescue tasks, food distribution, and physical and mental treatment (Fallah and Hosseini Nejad, 2020). The study has revealed that the disaster victims were satisfied with NGOs' actions but the governmental organisations were not able to coordinate well with them. This study has also highlighted that good and controlled communication is highly required to make any task effective.

**3.3 Community/Citizens:** Community/citizens appear as one major stakeholder in disaster relief management (Drennan and Morrissey, 2019). Citizens' engagement has proven to be crucial for faster recovery after any disaster. A study conducted in Australia and the United Kingdom (UK) highlighted that community resilience during disaster management is a multi-dimensional aspect (Coles and Buckle, 2004, Drennan and Morrissey, 2019). The community appears in different but equally important ways in response to disaster relief management. For example, any family or group of people appearing as a leading role in communicating with other schools/park areas to establish temporary shelter after a disaster. Another study on community responses to Hurricane Katrina demonstrated the importance of local knowledge, resources, and cooperative strategies in determining their survival and recovery (Patterson et al., 2010). In this study, It has been noted that the Jewish Family Service (JFS) compiled an Emergency Care Contact List, for which senior citizens could voluntarily pre-enrol. After the hurricane, JFS delivered services and assistance to seniors who are known to be potentially vulnerable. Similarly, the leadership of the Vietnamese community evacuated everyone who would leave and alerted Vietnamese communities in the surrounding areas. Another study conducted by Athe Australian emergency management doctrine recognised and emphasised the importance of community-led recovery in terms of enabling communities to take ownership of their recovery process and leverage the skills, capabilities and resources within an affected area, building their resilience through the recovery process a different study conducted by the International Federation of the Red Cross emphasises that local knowledge enables decision-makers to more effectively engage with affected populations and act as a mediator to both identify community needs, organise recovery efforts and build resilience. This research has suggested that local community leaders are more engaged in the community over the extended timeframe of disaster recovery (Drennan and Morrissey, 2019). In general, the community is often associated with social factors that are associated with self-motivated cooperation. The self-motivated community plays a positive role in disaster resilience.

**3.4 Leadership:** In disaster relief management, the role of the leader is extremely vital in bringing order to the state of calmness for disaster victims. Logically, leaders in disaster scenarios may be of appearing with different hierarchies and ranks, such as the head of the disaster central management team, and the local community leader where the disaster had occurred. In normal conditions, leadership decisions are made after a process of discussion in the coordination of experts' advice. However, leadership in a disaster scenario appears as applying the strategic tasks that incorporate all activities associated with disaster relief management. These roles of leaders in disaster scenarios are different from normal scenarios as under the disaster safety, support, evacuation, positivity and hope are among the top priorities (Mahmud et al., 2020). For example, after the Indian Ocean tsunami in 2004, the Indonesian President stated the tsunami was a national disaster and had the deployment of available resources for disaster relief. In the same disaster, the Indonesian Vice-President sent a high-ranking official from Jakarta to disaster relief tasks, while the President asked the international community for open emergency relief support. The president issued orders that allowed ease of access to international flights, visa requirements and exclusion from customs taxes for relief supplies (Mahmud et al., 2020). A report by the Stockholm International Peace Research Institute described that on the following day of the declaration, there was overwhelming international support for disaster resilience. A study from Japan Tsunami 2011 showed that community leaders had demonstrated 'active leadership' in the identification of objectives and identification of stakeholders and understanding of the sociocultural perspective for disaster relief management (Lin et al., 2017). In both studies, it has been observed that the active and timely role of leadership gives an effective direction for disaster resilience.

**3.5 Media**: Communication is also one of the major stakeholders in disaster relief management. Communication in a disaster scenario can be established using various ways such as television, newspaper, leaflets, SMS, audio/visible signals and the internet. Nowadays, different forms of social media such as Facebook, Twitter, Whatsapp Messenger, and Instagram, ranging from prompt messaging to social networking sites appear as effective tools for communication with citizens (Ahmed, 2011). Communication through mass media in a disaster scenario has the advantage of communicating with larger groups even in cases of a partial failure of the existing communication infrastructure (Nair, 2010). In many countries, overall disaster relief management has been released often without highlighting the importance of media and the right communication channel for disaster warnings or alerts (Nayak, 2012). But the fact is media plays a key role in motivating and allowing disaster victims to prepare for disaster resilience and reasonably apply themselves during and after the disaster. If the disaster warning or alert is made timely and accurately then it helps all responsible stakeholders to act towards effective disaster relief management. Media has its role in three stages including before, during, and after the occurrence of disasters. On the other hand, it has also been observed that media can also contribute to worsening the situation with cross-purpose disaster management when the message is not been communicated on time or the message is inaccurate (Soltani, 2015). To avoid such a situation, the governmental control mechanism is crucial. The control mechanism directs the media to reach the disaster victims with the right and timely messages.

## 4. Discussion and Conclusion

Natural disasters are inevitable and it is almost impossible to fully regain the damage caused by the disasters. However, it is possible to minimize the potential risk by developing effective disaster relief management such as preparing early disaster warning strategies, implementing relief items distribution to the disaster victims and helping in rehabilitation and post-disaster reduction. The role of government organisations is crucial as they establish the first stage of disaster relief management. However, it is important to note that the lack of knowledge on the disaster-affected region, affected individuals and groups, and geographical and cultural diversity at the governmental management level may have a negative impact on distribution relief management. To overcome the negative impact, during disaster scenarios, collaborative shared responsibility can be defined. The collaboration can be on the existing network or with any stakeholders that can contribute even if there has not been any no previous history of collaboration. Government organisations play a crucial role in developing a more stable and controlled collaborative shared responsibility by providing the needed resources and expertise. Despite evolving consent for collaborative shared responsibility in disaster management, there is still no clear arrangement on what and how the responsibility can be shared (Busch and Givens, 2013). But, the effective engagement and management of multiple stakeholders in coordination with government organisations is a key element of the disaster recovery process. Another challenge for shared responsibility is to maintain a sense of accountability. Frequent communication, knowledge sharing, information exchange and stating clearly the expectations from each stakeholder can undoubtedly improve the shared responsibility as these lead to converging the disaster relief management objectives.

The roles of NGOs in disaster relief management have been increasing due to the rising challenges of disaster occurrence and its uncertainty. In a disaster scenario, the major role of the NGO is to establish a strong relationship among the government organisations, leadership and community for the effective mitigation of disaster resilience. Collaboration between NOGs with government organisations and other sectors complements the relief tasks with their expertise and resources. More often, it has been observed that there are many NGOs that have a direct link to disaster regions and disaster victims as they have worked more closely with them in normal scenarios. In addition, the NGOs act promptly in disaster scenarios in a flexible way to support the other stakeholders for disaster resilience. Leaders in disaster relief management are another key stakeholder as they have the ability of decisions making manage good relationships with disaster victims. To bring higher effectiveness in disaster relief management, leaders need to familiarise themselves with the disaster region, and cultural and geographical diversity and accommodate that knowledge in their decision-making processes. In disaster relief management, leaders often carry out leadership roles of coordination that reduce susceptibility and minimise the loss of life and property. Additionally, effective leadership in a disaster situation is when the leader can comprehensibly coordinate different agencies to minimise the disaster's impact.

Engagement of the community is one of the most important aspects of disaster that needs to be considered in every disaster relief management. Active participation of citizens in minimising the disaster impact and recovery helps to successfully implement disaster resilience plans. Therefore, disaster relief management plans, involving disaster-affected citizens needs to be considered as part of the sustainable action plan. It has been observed that not every community is equal and beneficial therefore the strengths and weaknesses of the community must be taken into account in disaster relief management. The engagement of the community has a positive impact as more often the well-functioning community organisations have the faith of their members and hold the moral authority to adopt cooperative behaviour and teamwork at the local level which is harder for governmental organisations. The local communities also have strong abilities to assess individual requirements and distribute relief items efficiently and equitably. Because of this local-level knowledge, it is important to include local communities in disaster relief management and establish collaborated shared responsibility between communities and government organisations. The media is treated as another key stakeholder in disaster relief management as the media can help the timely broadcast information to the wider community in a short time. Media can increase the speed of response in disasters with the proper spreading of information which eventually helps in speeding up the process of disaster resilience. The media also can communicate in both directions by articulating the disaster victim's emotions to the government organisations. However, there are a few issues such as having incomplete and untimely messages, conflicting information from different media, and setting accountability in social media are the major challenges. The control mechanism of government organisations can minimise these negative impacts on media communication.

In this paper, we presented a framework with collaborative shared responsibility for disaster relief management. The framework presents the five key stakeholders and their responsibility in disaster resilience. The framework will have a positive impact on the operational efficiency of disaster relief management and also can respond faster to different stakeholders after any disaster. Tough, the scope and nature of collaboration and responsibility may vary by disaster types, impacts, and cultural and geographical diversity. In each country, such as Nepal, there has been wider diversity in culture and geography. These diversities bring a high risk of uncontrolled relief management. The presented framework guides the establishment of the right communication and hence the precise control mechanism to make the relief management task more effective. In summary, the shared responsibility makes disaster relief management more flexible in terms of effective relief item distribution. Operationally, shared responsibility enables government organisations to manage the resources rapidly, making the relief distribution system more responsive even in scenarios with cultural and geographical diversity.

## References

AHMED, A. 2011. Use of social media in disaster management.

ATKINSON, C. & CURNIN, S. 2020. Sharing responsibility in disaster management policy. *Progress in disaster science,* 7**,** 100122.

BOSHER, L., DAINTY, A., CARRILLO, P., GLASS AND, J. & PRICE, A. 2007. Integrating disaster risk management into construction: a UK perspective. *Building research and information,* 35**,** 163-177.

BUSCH, N. E. & GIVENS, A. D. 2013. Achieving resilience in disaster management: The role of public-private partnerships. *Journal of strategic security,* 6**,** 1-19.

CHEONG, S.-M. 2011. The role of government in disaster management: the case of the Hebei Spirit oil spill compensation. *Environment and Planning C: Government and Policy,* 29**,** 1073-1086.

COLES, E. & BUCKLE, P. 2004. Developing community resilience as a foundation for effective disaster recovery. *Australian Journal of Emergency Management, The,* 19**,** 6-15.

CROSWELLER, M. & TSCHAKERT, P. 2021. Disaster management and the need for a reinstated social contract of shared responsibility. *International Journal of Disaster Risk Reduction,* 63**,** 102440.

DRENNAN, L. & MORRISSEY, L. 2019. Resilience policy in practice—Surveying the role of community based organisations in local disaster management. *Local Government Studies,* 45**,** 328-349.

FALLAH, S. & HOSSEINI NEJAD, J. 2020. The Role of Non-Governmental Organizations in Disaster Management: A Case study of Bam Earthquake, Iran. *Journal of Disaster and Emergency Research,* 1**,** 43-50.

GUERDAN, B. R. 2009. Disaster preparedness and disaster management. *Am J Clin Med,* 6**,** 32-40.

HOYOS, M. C., MORALES, R. S. & AKHAVAN-TABATABAEI, R. 2015. OR models with stochastic components in disaster operations management: A literature survey. *Computers & Industrial Engineering,* 82**,** 183-197.

LASSA, J. A. 2018. Roles of non-government organizations in disaster risk reduction. *Oxford Research Encyclopedia of Natural Hazard Science.*

LIN, Y., KELEMEN, M. & KIYOMIYA, T. 2017. The role of community leadership in disaster recovery projects: Tsunami lessons from Japan. *International Journal of Project Management,* 35**,** 913-924.

MAHMUD, A., MOHAMMAD, Z. & ABDULLAH, K. A. 2020. Leadership in disaster management: Theory versus reality. *Journal of Clinical and Health Science,* 5**,** 4-11.

MEDEL, K., KOUSAR, R. & MASOOD, T. 2020. A collaboration–resilience framework for disaster management supply networks: a case study of the Philippines. *Journal of Humanitarian Logistics and Supply Chain Management.*

NAIR, P. 2010. Role of media in disaster management. *Mass Communicator,* 4**,** 36-40.

NAYAK, K. 2012. Role of media in disaster reduction. *Odisha Review,* 2**,** 135-147.

PATTERSON, O., WEIL, F. & PATEL, K. 2010. The role of community in disaster response: conceptual models. *Population Research and Policy Review,* 29**,** 127-141.

RAWLS, C. G. & TURNQUIST, M. A. 2010. Pre-positioning of emergency supplies for disaster response. *Transportation research part B: Methodological,* 44**,** 521-534.

REDUCTION, I. S. F. D. 2004. Living with risk: A global review of disaster reduction initiatives.

SOLTANI, F. 2015. Mass media and its role in increasing society's involvement in disaster management. *International Journal of Health System and Disaster Management,* 3**,** 12.